\documentclass[aps,pre,twocolumn,showpacs,showkeys,a4paper]{revtex4-1}
 \usepackage{graphicx}
 \usepackage{amsmath}
 \usepackage{amssymb}
 \usepackage{amscd}
 \usepackage{color}

\begin{document} 

 \title{Dynamics of a polymer in an active and viscoelastic bath}
 \author{Hans Vandebroek$^{(1)}$ and Carlo Vanderzande$^{(1,2)}$}
 \affiliation{(1) Faculty of Sciences, Hasselt University, 3590 Diepenbeek, Belgium\\
  (2) Instituut Theoretische Fysica, Katholieke Universiteit Leuven, 3001
  Heverlee, Belgium}


 \begin{abstract} 
We study the dynamics of an ideal polymer chain in a viscoelastic medium and in the presence of active forces. The motion of the centre of mass and of individual monomers is calculated. On time scales that are comparable to the persistence time of the active forces, monomers can move superdiffusively while on larger time scales subdiffusive behaviour occurs. The difference between this subdiffusion and that in absence of active forces is quantified. We show  that  the polymer swells in response to active processes and determine how this swelling depends on the viscoelastic properties of the environment. Our results are compared to recent experiments on the motion of chromosomal loci in bacteria. 
 \end{abstract}

\maketitle 
\section{Introduction}
The dynamics of a polymer in a viscous solvent in equilibrium is well understood \cite{Rouse53,Doi86}. 
Starting from the exactly solvable Rouse model \cite{Rouse53}, which neglects excluded volume and hydrodynamic interactions, it was found that the centre of mass of the polymer diffuses with a diffusion constant $D$ that is inversely proportional to the number of monomers $N$. The time for the polymer to diffuse over its own radius of gyration $R_G (\sim N^{1/2})$ then introduces a typical time scale, called {\it Rouse timescale}, $\tau_R$, which scales with $N$ as $\tau_R \sim R_G^2/D\sim N^2$. Individual monomers subdiffuse for times smaller then $\tau_R$, but follow the diffusion of the centre of mass after $\tau_R$ \cite{Doi86}. Going beyond the Rouse model, the effects of excluded volume and hydrodynamic interactions can be incorporated using scaling theories \cite{deGennes79}, simulations and exactly solvable models \cite{Vanderzande98}.

Much less is known about the dynamics of a polymer in the crowded \cite{McGuffee10} and nonequilibrium cellular environment. The crowdedness introduces viscoelastic behaviour \cite{Guigas07} with a long term memory while  the action of molecular motors and other ATP-driven, active, processes puts the cell out of equilibrium. 

Recently, several experimental and theoretical studies have investigated the role of active processes on biopolymer dynamics. 
On a coarse grained scale, activity can be seen as an extra source of randomness in addition to that due to thermal motion. This is reflected in
a random motion of tracer particles (and small biopolymers like proteins) that is much enhanced in comparison with  thermal Brownian motion \cite{Brangwynne08,Lau03, Robert10, Fakhri14}.  
In other experiments, the internal dynamics of a polymer was found to be influenced by the presence of active forces. 
We mention the effect on the bending dynamics of microtubuli  \cite{Brangwynne08b}. More relevant for the present paper are studies in which it was found that the motion of chromosomal loci in simple organisms like bacteria and yeast are sensitive to active forces \cite{Weber12, Javer12}.  For example, Weber and collaborators \cite{Weber12} found that after addition of chemicals that inhibit ATP-synthesis, the diffusion constant of chromosomal loci decreased by 49$\%$. Also measurements of chromatin dynamics in eukaryotes show evidence for an important role played by ATP-dependent processes \cite{Zidovska13}.

In the theoretical description of the motion of chromosomal loci, it came as somewhat of a surprise that again the simple Rouse model turned out to be relevant \cite{Liu15}.
Indeed, it was recently argued to give a good description of this motion, both between \cite{Weber10} and during chromosomal segregation \cite{Lampo15}. The reason for this may be the action of topoisomerases and related enzymes which cross chromosome strands and thus make the bacterial chromosome a phantom chain \cite{Liu15}. However, other models have been introduced to describe the motion of bacterial chromosomes, for example, in terms of self-adhesion of monomers \cite{Scolari15}. None of these models did however explicitly investigate the effects of active forces.

In the present paper, we extend recent work on the Rouse chain in a {\it viscoelastic} medium \cite{Weber10b, Vandebroek14, Lampo15} by including active forces. The advantage of our model is that it is exactly solvable and hence like the original Rouse model can be used as a benchmark for studies of more realistic models that include physical effects like self-avoidance, bending rigidity and so on. Some of our results (superdiffusion, polymer swelling) have indeed been recently seen in simulations \cite{Ghosh14,Kaiser14} of such models. However, all the numerical work known to us is for the {\it viscous} regime. Our work is the first to include both effects of viscoelasticity and active forces, two ingredients that are necessary for a proper modelling of  a cellular environment.

\section{Model}  
The Rouse model \cite{Rouse53,Doi86} is the starting point in all discussions of polymer dynamics. It models a long polymer chain as a set of $N$ beads (monomers) connected by harmonic springs. It is important to point out that such a description of a real polymer is only appropriate at a coarse grained level, i.e. at length scales above the persistence length. Hence, our model will only be relevant for long biopolymers (like chromosomes) and also other physical properties, like crowdedness, viscoelasticity and active forces, will be described at this coarse grained level.

\begin{figure}[h]
\centering
\includegraphics[width=8cm]{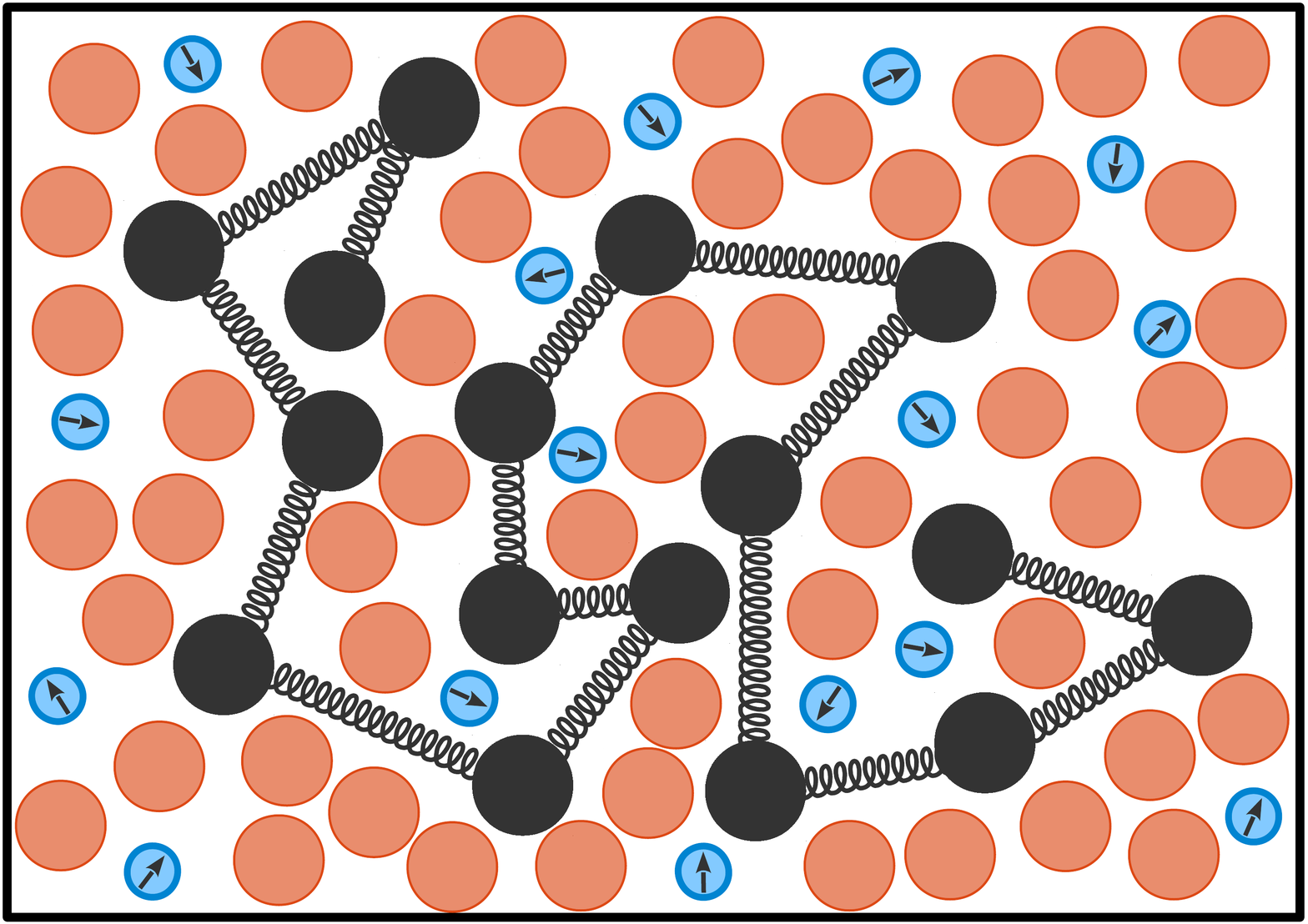}
\caption{(Color online) Cartoon of our model : a large polymer (black spheres connected with springs) in a dense crowded cellular environment (orange spheres represent other smaller biopolymers like proteins, RNA, ...) that is brought out of equilibrium by active processes (blue spheres with arrows represent active particles that for a time of order $\tau_A$ move in a fixed direction).  }
\label{Fig1}
\end{figure}

We denote by $\vec{R}_n(t)$ the position of the $n$-th monomer ($n=0,1,\ldots,N-1$) at time $t$. Let us discuss the various forces acting on this monomer (for a schematic representation of our model, see Fig. \ref{Fig1}). 

In the Rouse model, the monomers are connected by springs with spring constant $k$. In equilibrium, the average squared distance $b^2$ between two monomers then follows from the equipartition theorem and equals $b^2=3k_B T /k$ where $k_B$ is Boltzmann's constant and $T$ temperature. A Rouse chain gives a description of a semiflexible polymer on length scales where $b$ corresponds to the Kuhn length \cite{Rubinstein03}.  We will apply our model to the bacterial chromosome, for which the Kuhn length is of the order of $100$ nm or $300$ base pairs (bp) \cite{Lampo15}. Since the chromosome of {\it E. Coli} has approximately $4.5 \times 10^6$ bp,  the relevant value of $N$ is of the order $10^4$. 

The cellular environment is a crowded material \cite{Guigas07,Luby87, Fabry01} in which tracer particles (nano particles, mRNA, ...) were found to subdiffusive (for a recent review, see \cite{Metzler13}). While originally the precise explanation of this behaviour was unclear, and various mechanisms where proposed, recent experimental evidence shows that viscoelasticity and its mathematical description based on a generalized fractional Langevin equation best describes the observed behaviour \cite{Weiss13}.  The friction force $\vec{F}_n(t)$ on a monomer in a viscoelastic medium has memory and is commonly described in terms of a power law kernel $K(t)=(2-\alpha)(1-\alpha) t^{-\alpha}$
\begin{eqnarray}
\vec{F}_n(t) = -\gamma \int_0^t d\tau K(t-\tau) \frac{d\vec{R}_n(\tau)}{dt} \label{1}
\end{eqnarray}
Viscous behaviour is recovered for $\alpha=1$ (for which $K(t) \to \delta(t)$) while for $\alpha=0$ we get elastic behaviour. For $0 < \alpha < 1$, we have the viscoelastic situation, intermediate between elastic and viscous response. Estimates of $\alpha$ range from $\alpha \approx 0.7$ for {\it E. Coli} \cite{Weber10} to $\alpha \approx 0.2$ in the cytoplasm \cite{Fakhri14} of eukaryotes. An expression for the friction like Eq.~\eqref{1} breaks down on short timescales of the order of the molecular collision time scale. In bacterial cells, an upper limit to memory effects is put by the cellular lifetime. 

The random thermal force $\vec{\xi}_{T,n}(t)$ acting on the $n$-th monomer is given by a Gaussian random variable, with average zero and a correlation that is coupled to the kernel $K(t)$ by the (second) fluctuation-dissipation theorem \cite{Zwanzig01}
\begin{eqnarray}
\langle \vec{\xi}_{T,n}(t)\cdot \vec{\xi}_{T,m}(t') \rangle =  3 \gamma k_B T  K(|t-t'|)\ \delta_{n,m}
\label{2}
\end{eqnarray}
With the choice of the power law kernel, $\vec{\xi}_{T,n}(t)$ becomes fractional Gaussian noise \cite{Mandelbrot68}.

Much less is known about the precise form of the active forces $\vec{\xi}_{A,n}(t)$. In fact, the precise characterisation of their statistical properties in living cells is a topic of current research \cite{Robert10,Levine09,Bohec13,Fodor14,Fodor15}. Ultimately, the activity is due to active 'particles' (molecular motors, or other active proteins) that consume energy to generate motion and associated dissipation. In a simple model (see \cite{Marchetti15} for a review), self-propelled particles (SPPs) move with constant velocity in a direction $\vec{e}$ that is subject to rotational Brownian dynamics. This leads to a typical motion in which the autocorrelation of $\vec{e}$ decays exponentially. In \cite{Kaiser14}, the dynamics of a polymer in the presence of such SPPs is studied. The polymer and SPPs interact through a truncated Lennard-Jones potential. On a more coarse grained level, this type of interaction leads to a random force on the monomers. The force has an exponential correlation with a timescale $\tau_A$, which can be interpreted as the typical time during which the SPPs move in a straight line. 

The modelling of active processes through this type of random forcing is quite common in the literature
and was for example used in \cite{Ghosh14} in simulations of active semiflexible polymers. We will follow this approach since our model, as already stressed above, is defined on a coarse grained scale.
In conclusion then, we assume that the active force on the $n$-th monomer, $\vec{\xi}_{A,n}(t)$, is a Gaussian random variable with average zero and a correlation given by
\begin{eqnarray}
\langle \vec{\xi}_{A,n}(t)\cdot \vec{\xi}_{A,m}(t') \rangle =  3 C  \exp\left(-|t-t'| /\tau_A\right) \delta_{n,m}
\label{3}
\end{eqnarray}
Here $C$  characterises the strength of the active noise. We do not include possible spatial correlations in order to keep the model soluble. Moreover, little experimental insight on such correlations is available.

Putting everything together and neglecting inertial terms,  the equation of motion of the $n$-th monomer is  the overdamped generalized Langevin equation
\begin{eqnarray}
\vec{F}_n(t) &-&k \left(2 \vec{R}_n(t) - \vec{R}_{n+1}(t) - \vec{R}_{n-1}(t) \right) \nonumber \\
&+& \vec{\xi}_{T,n}(t) + \vec{\xi}_{A,n}(t) H(t)=0
\label{0}
\end{eqnarray}
where $H(t)$ is the Heaviside function. It is important to point out that the active forces, since they are not related to the friction kernel, put the system out of equilibrium. Starting in equilibrium at $t=0$, the solution to equation (\ref{0}) therefore gives the response of the polymer to, for example, the addition of ATP at $t=0$. After a long time, the polymer will evolve to a new, nonequilibrium steady state. In the next section we will calculate both the transient and steady state behaviour of the polymer after activation of the active forces. 

\section{Results} 
The techniques to solve the set of equations (\ref{0}) with appropriate boundary conditions are standard \cite{Doi86}. The details are given in the supplemental material to this paper \cite{sup}. Here, we discuss only the results.

In the dynamics of the polymer two relevant time scales occur. The first is $\tau_A$, the second is the Rouse time $\tau_R$, which in the viscoelastic medium equals $\left[ \Gamma(3-\alpha) \gamma b^2 N^2/3 k_B T \pi^2 \right]^{1/\alpha}$ \cite{Weber10b, Vandebroek14}. Notice that for the bacterial chromosome, using the values of $b, N$ and $\alpha$ quoted above, $\tau_R$ can become quite large, indeed larger than the duration of the cell cycle. The persistence time of active processes should be much smaller then the cellular lifetime, so that when applying our model to the bacterial chromosome, the two timescales are separated, $\tau_A \ll \tau_R$. Experiments on chromosomal loci are done on timescales of $0.1$ seconds to minutes, i.e. in the regime $\tau_A \sim t \ll \tau_R$. 

\begin{figure}
\vspace{0.5cm}
\includegraphics[width=9cm]{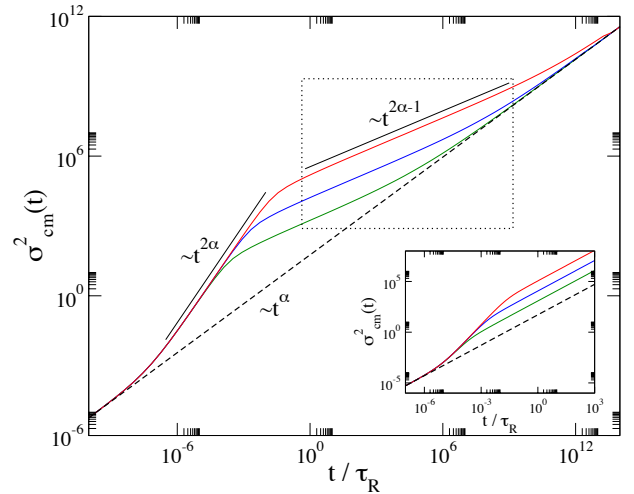}
\caption{\label{Fig2} (Color online) Log-log plot of squared distance travelled by  the centre of mass as a function of $t/\tau_R$ for a Rouse chain in a viscoelastic medium ($\alpha=0.7, k_BT=\gamma=1$) in the presence of active forces with $C=100$ and $ \tau_A/\tau_R=10^{-2}, 10^{-3}, 10^{-4}$ (full lines, top to bottom) compared to that without active forces (dashed line). The inset shows the same for the viscous case, $\alpha=1$. Results are for $N=256$. }
\end{figure}

We first discuss the motion of the centre of mass $\vec{R}_{cm}(t)$.
It was found earlier that in a viscoelastic medium but in absence of active forces \cite{Weber10}, the centre of mass $\vec{R}_{cm}(t)$ performs a subdiffusion
\begin{eqnarray}
\sigma^2_{cm}(t)= \Big\langle \left( \vec{R}_{cm} (t) - \vec{R}_{cm} (0) \right)^2 \Big\rangle = \frac{6 k_B T}{\gamma N \alpha G_\alpha} t^\alpha
\label{4}
\end{eqnarray}
where $G_\alpha=\Gamma(\alpha) \Gamma(3-\alpha)$. To this subdiffusion an extra term is added when the active forces are turned on. It equals
\begin{eqnarray}
\frac{6 C \tau_A^{2 \alpha}}{N \gamma^2 G_\alpha^2}\ \int_0^{t/\tau_A} dy\ e^y y^{\alpha-1} \Gamma(\alpha; y, t/\tau_A)
\label{5}
\end{eqnarray}
where $\Gamma(\alpha; y,x)$ is a difference of two incomplete gamma-functions. Firstly, observe that as in the standard Rouse chain, the (generalized) diffusion constant remains inversely proportional to $N$. More interestingly, it can be shown that for $t \ll \tau_A$, (\ref{5}) goes as $t^{2 \alpha}$, i.e. is superdiffusive if $\alpha > 1/2$. On the other hand, for $t \gg \tau_A$, (\ref{5}) evolves as $t^{2 \alpha-1}$, i.e. slower than (\ref{4}), so that asymptotically in time, $\sigma^2_{cm}(t)  \sim t^\alpha$. The resulting behaviour is shown in Fig. \ref{Fig2} for $\alpha=0.7, C=100$ and various values of $\tau_A$ \cite{Remark}. As the figure shows (boxed region), if $C\tau_A$ is large compared to $\gamma k_B T$, over several orders of magnitude in time, the centre of mass subdiffuses with an exponent $2 \alpha - 1$, whereas without active forces, the exponent would be $\alpha$. If this time regime corresponds to the experimental one, not taking into account active forces could lead to a wrong estimate of the exponent $\alpha$, 
and hence to a wrong characterisation of the rheological properties of the environment. Results like that of Fig. \ref{Fig2} are under the assumption that our model is realistic for all $t$. When comparing with experimental data, one has to take into account that, as already discussed in the previous section, the model will break down on small and large time scales. Therefore in the cellular context, it is possible  that only the superdiffusive behaviour is observed. 

For the viscous case, $2 \alpha - 1 = \alpha$, so that after an initial regime of ballistic motion $\sigma^2_{cm}(t) \sim t^2$, the polymer performs ordinary diffusion but with a diffusion constant that is enhanced by a factor $1+ C \tau_A/\gamma k_B T $ (see inset of Fig. \ref{Fig2}).

\begin{figure}
\includegraphics[width=9cm]{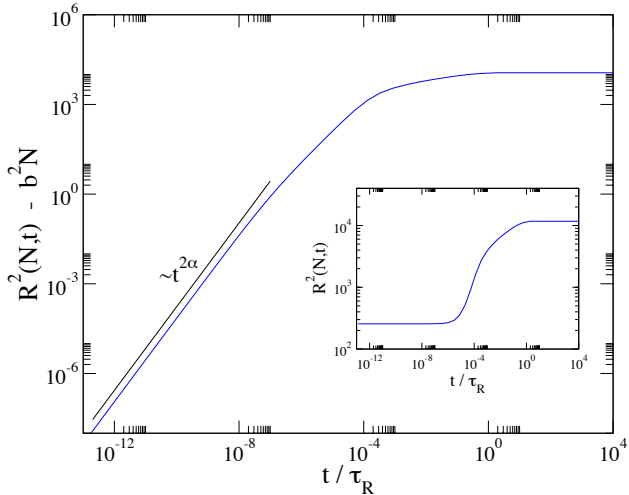}
\caption{\label{Fig3} (Color online) Squared end-to-end distance of the polymer as a function of time. In the main figure (inset), the equilibrium length is (not) subtracted. The straight line has slope $1.4$. ($\tau_A/\tau_R=10^{-4}, C=100, \alpha=0.7, N=256$)}
\end{figure}

A second global quantity is the end-to-end vector $\vec{P}(t) = \vec{R}_0(t)-\vec{R}_{N-1}(t)$.  Its averaged squared length, $R^2(N,t)=\langle \vec{P}(t) \cdot \vec{P}(t) \rangle$, measures the size squared of the polymer and equals in equilibrium, both in viscous and viscoelastic media, $b^2 N$. In response to active forces, $R^2(N,t)$ gets an additional term which equals
\begin{widetext}
\begin{eqnarray}
\frac{24 C}{N \gamma^2 \Gamma^2(3-\alpha)} \sum_{p=1,odd}^{N-1} \int_0^t d\tau \int_0^t  d\tau'
e^{-\frac{|\tau-\tau'|}{\tau_A}} \tau^{\alpha-1} \tau'^{\alpha-1} E_{\alpha,\alpha} \left(-\left(\frac{\tau}{\tau_{p}}\right)^\alpha\right) E_{\alpha,\alpha} \left(-\left(\frac{\tau'}{\tau_{p}}\right)^\alpha\right) \label{7}
\end{eqnarray}
\end{widetext}
Here $E_{\alpha,\beta}(z)$ is the generalized Mittag-Leffler function \cite{Haubold11}, and $\tau_{p}=\tau_R/p^{2/\alpha}$.  Since (\ref{7}) is positive, we conclude that active forces swell the polymer. It can be easily shown that initially this swelling grows proportional to $t^{2 \alpha}$ after which $R^2(N,t)$ saturates. In Fig. \ref{Fig3}, we show the results of a numerical evaluation of (\ref{7}) for a polymer with $N=256$ in a medium with $\alpha=0.7$. 

The value at which $R^2(N,t)$ saturates, i.e. the squared size of the polymer in the new, nonequilibrium steady state, has an interesting $N$-dependence. For the viscous case, $\alpha=1$, the integrals in (\ref{7}) can be easily calculated. In this way, it is found that the swelling of the polymer is proportional to $N$ so that
\begin{eqnarray}
R_{ne}^2(N) \equiv \lim_{t \to \infty} R^2(N,t) = \left(1 + \frac{C\tau_A}{\gamma k_B T}\right) b^2 N
\label{8}
\end{eqnarray}
In fact this result is exactly what can be expected from a recent extension of the equipartition theorem to harmonic oscillators in viscous, active media \cite{Maggi14}. In that reference, the average potential energy of a harmonic oscillator in a viscous, active bath is calculated and compared with experiments. Using the results of that paper, and the fact that the Rouse chain consists of $N$ independent harmonic oscillators one can also derive (\ref{8}). 

\begin{figure}[here]
\vspace{0.5cm}
\includegraphics[width=9cm]{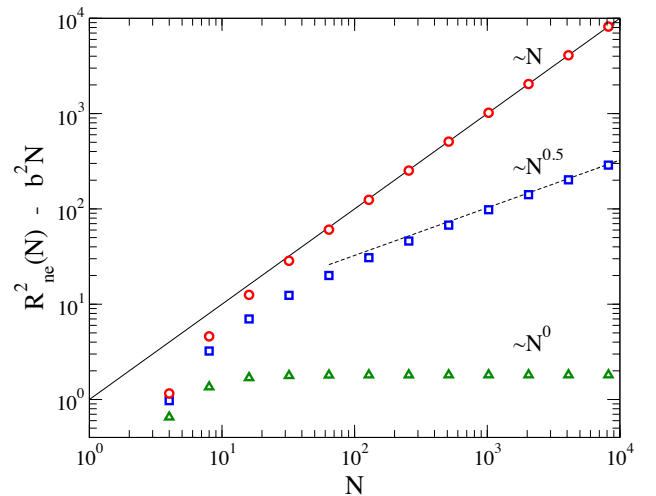}
\caption{\label{Fig4} (Color online) Difference between squared end-to-end distance of the polymer in the nonequilibrium steady state and in equilibrium as a function of $N$ for $\alpha=1, 0.8$ and $0.4$ (top to bottom). The symbols are the result of a numerical evaluation of (\ref{7}) for $t \to \infty$. The full line is a plot of (\ref{8}), the dashed line of (\ref{9}). }
\end{figure}

In the viscoelastic case, the situation is more complicated. It is possible to determine the leading behaviour of the integrals in (\ref{7}) for  $N \gg 1$ (and $t \to \infty$). In this way it is found that for $2/3 < \alpha < 1$
\begin{eqnarray}
R_{ne}^2(N) = b^2 N +  \frac{48 C \tau_A (4\pi^2 k)^{1/\alpha-2}}{(\gamma \Gamma(3-\alpha))^{1/\alpha}} f(\alpha) N^{3 - \frac{2}{\alpha}} 
\label{9}
\end{eqnarray}
where 
\begin{eqnarray*}
f(\alpha)=\zeta(4-2/\alpha,1/2) \int_0^\infty dx\ x^{2\alpha-2} E^2_{\alpha,\alpha} \left(-x^\alpha\right)  
\end{eqnarray*}
and $\zeta(x,y)$ is the Hurwitz zeta function. 
In Fig. \ref{Fig4} we show the results of a numerical evaluation of (\ref{7}) for $\alpha=0.8$ together with the asymptotic behaviour (\ref{9}) (dashed line). For $\alpha < 2/3$, numerical integration of (\ref{7}) indicates that the swelling of the polymer approaches a constant as $N$ increases (Fig \ref{Fig4}). From our calculations, we see that in an active environment, large polymers are orders of magnitude more compact in a viscoelastic medium with small $\alpha$. This observation could be of relevance for storing a large chromosome in a small cell. 

Finally, we turn to the motion of the individual monomers. From an experimental point of view, this quantity is the most interesting one, since it can be determined using fluorescence techniques \cite{Weber12,Javer12,Kuwada13,Javer14}. We present results for the middle monomer, but the behaviour of other monomers is qualitatively the same (see supplemental material for the general case \cite{sup}).

In absence of active forces, one has \cite{Weber10}
\begin{multline}
\sigma^2_m(t) = \Big\langle \left( \vec{R}_{N/2}(t) - \vec{R}_{N/2}(0)\right)^2 \Big\rangle = \frac{6 k_B T}{\gamma N\alpha G_\alpha} t^\alpha
 \\
+ \frac{4b^2 N}{\pi^2} \sum_{p=2,even}^{N-1}\frac{1}{p^2} \left[ 1 - E_{\alpha,1}\left(-\left(\frac{t}{\tau_{p}}\right)^\alpha\right)\right] \label{10}
\end{multline}

For $t < \tau_R$, the monomer is found to subdiffusive with an exponent $\alpha/2$ whereas on time scales larger than $\tau_R$, the monomer follows the motion of the centre of mass, hence subdiffuses with an exponent $\alpha$. Since for the cytoplasm of {\it E. Coli}, $\alpha$ equals approximately $0.7$,  the observed exponent of the subdiffusion of the chromosomal loci, $0.39 \pm 0.04$ \cite{Weber10, Javer12}, is consistent with $\alpha/2$ within the experimental error. 

In the presence of active forces two terms have to be added to (\ref{10}). The first one is the same as for the centre of mass (\ref{5}), the second equals $1/4$ of (\ref{7}) (but now with the sum over the even modes). 

\begin{figure}
\vspace{0.5cm}
\includegraphics[width=9cm]{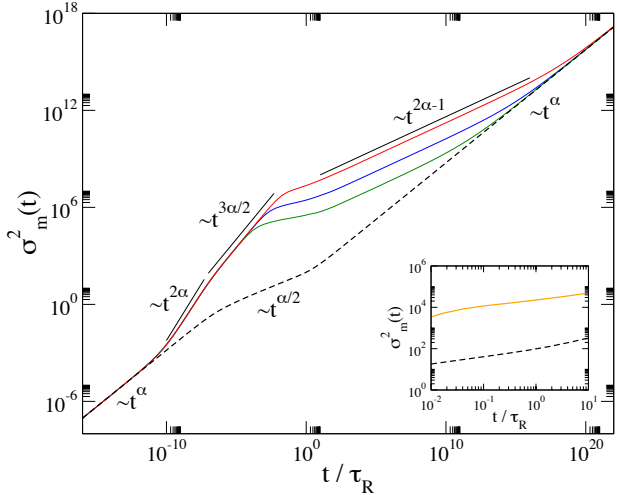}
\caption{\label{Fig5} (Color online) Log-log plot of $\sigma^2_m(t)$ as a function of time in presence of active forces (full lines) for $\alpha=0.7, k_BT=\gamma=1, C=10^4, N=256$ and $\tau_A/\tau_R=10^{-2}, 10^{-3}, 10^{-4}$ (top to bottom). The dashed line is the same quantity in absence of active forces. The inset shows that the results for a smaller time window  in absence and presence of active forces ($C=1, \tau_A/\tau_R=10^{-2}$) run parallel.}
\end{figure}

Analysing the resulting behaviour, we find that as was the case for the centre of mass, the short time response of the monomers to active forces is superdiffusive (at least when $\alpha>1/2$). This behaviour, where $\sigma^2_m(t) \sim t^{2\alpha}$,  holds for $t \le  \tau_A$. Very recently superdiffusive motion of chromosomal loci has indeed been observed \cite{Javer14}. It is not clear whether these {\it rapid chromosomal movements} are due to active processes or to stress relaxation. Our results however quantify better the response to active forces and could therefore help in discriminating the real origin of the observed motion. 

For $t \gg \tau_R$ the monomers follow the $t^\alpha$ subdiffusion of the centre of mass. The behaviour in the intermediate time regime, $\tau_A \ll  t \ll \tau_R$,  is more complicated and contains terms proportional to $t^{\alpha/2}$ (coming form (\ref{10})), $t^{2 \alpha-1}$ (from the centre of mass), $t^{3\alpha/2}$ and $t^{3 \alpha/2-1}$. No simple power law behaviour emerges. A plot of the full expression for $\sigma_m^2(t)$ (Fig. \ref{Fig5}) for $C$ large shows that the initial superdiffusion is followed by a $t^{3\alpha/2}$ behaviour, after which there is a large time window in which the $t^{2 \alpha-1}$ term dominates. The crossover time to the $t^\alpha$-regime is very large, and may not be observable for chromosomal loci. 
For certain values of $C$ and $\tau_A$, there is a time regime (see inset of Fig. \ref{Fig5}) where, on a log-log scale, the graphs of $\sigma_m^2(t)$ in presence and absence of active forces run parallel.  This scenario resembles, at least qualitatively, the experimental one, where after inhibition of ATP synthesis with
sodium azide and 2-deoxyglucose, the exponent of the loci's subdiffusion hardly changed but the diffusion coefficient decreased \cite{Weber12}.
\section{Conclusions}
In summary, we have studied the behaviour of a long ideal polymer chain in a viscoelastic and active bath. We have formulated and solved a nonequilibrium version of the Rouse model, a model that in equilibrium forms the starting point of the theory of polymer dynamics. The results obtained are interesting for nonequilibrium statistical mechanics (power law response to active forces, equipartition theorem out of equilibrium) and polymer physics (swelling of polymer in an active bath). 
Moreover, our results show a qualitative similarity with experimental results on bacterial chromosomes. We mention the regime with superdiffusive motion of monomers and the fact that in certain time frames the exponent of the subdiffusive motion is almost independent of the presence/absence of active forces. 

While the Rouse model has been claimed to describe well several properties of bacterial chromosomes \cite{Weber10,Lampo15}, it is clear that such a description cannot be valid for all time and length scales and that effects of self-avoidance, bending rigidity, topology and so on have to be taken into account. This is even more true for chromatin for which modified Rouse models have been introduced that take into account bending rigidity \cite{Ghosh14} or long range interactions \cite{Amitai13}. The motion of chromosomal loci was also investigated on the basis of scaling arguments and computer simulations \cite{Tamm15} in the fractal globule model, which shares topological properties with chromatin \cite{Halverson14}. 
To investigate these issues further, we have recently developed an algorithm that allows to simulate particles and polymers in a viscoelastic medium subject to correlated noises coming both from thermal forces and active forces. With that algorithm it becomes possible to include effects of self-avoidance and bending rigidity and to see how the behaviour found here for an ideal chain is modified. The results will be published elsewhere \cite{Vandebroek16}. \\
\ \\

{\bf Acknowledgement} We thank M. Baiesi, M. Cosentino Lagomarsino and A.L. Stella for useful comments on the manuscript. We also thank E. Carlon and T. Sakaue for interesting discussions.


\begin{thebibliography}{99}
\bibitem{Remark}
In this and other figures, integrals like that appearing in (\ref{5}) are evaluated numerically using Mathematica.
\end{thebibliography}
\end{document}